%% file: main_file.tex
\documentclass[twocolumn,5p,times,final]{elsarticle}
\usepackage{pgfplots}
\pgfplotsset{compat=1.12}
\usepackage{graphicx}
\usepackage{amssymb, amsmath, amsfonts, amsthm}
\usepackage{multirow}
\usepackage{bm}
\usepackage{flushend}
\usepackage{booktabs}
\usepackage{colortbl}
\usepackage{enumitem}
\usepackage{multirow}
\usepackage{soul}
\usepackage{mathtools}
\input{JOURNAL/FIGURES/Style.tex}

\newcommand{\etal}{\textit{et al.}}
\journal{ }
\begin{document}
\begin{frontmatter}
\title{Network-Reconfiguration-Aware Power Oscillation Damping Controller for \\ Newly Commissioned Converter-Interfaced Power Plants
\tnoteref{projects_info}}
\author[inst1]{Njegos Jankovic\corref{cor1}}
\ead{njegos.jankovic@imdea.org}
\author[inst1]{Javier Roldan-Perez}
\author[inst1]{Milan Prodanovic}
\author[inst2]{Jon Are Suul}
\author[inst2]{Salvatore D'Arco}
\author[inst3]{Luis Rouco Rodriguez}
\affiliation[inst1]{
organization={Electrical Systems Unit, Imdea Energy},
addressline={Av. Ramon de La Sagra, 3},
city={Mostoles},
postcode={28935},
country={Spain}}
\affiliation[inst2]{
organization={Energy Systems, SINTEF Energy},
addressline={Strindvegen 4},
city={Trondheim},
postcode={7010},
country={Norway}}
\affiliation[inst3]{
organization={Instituto de Investigacion Tecnologica, Icai Universidad Pontificia Comillas},
addressline={Calle de Alberto Aguilera, 25},
city={Madrid},
postcode={28015},
country={Spain}
}
\cortext[cor1]{Corresponding author.}
\tnotetext[projects_info]{
This work has been financed by the following research projects: PROMINT (P2018/EMT4366), SOLARFLESS (TED2021-132854A-I00) and Juan de la Cierva Incorporaci\'{o}n program (IJC2019-042342-I).
The work of Njegos Jankovic was supported by a PhD Collaboration Agreement between Comillas Pontifical University and IMDEA Energy Institute.
}
\begin{abstract}
In recent years, transmission system operators have started requesting converter-interfaced generators~(CIGs) to participate in grid services such as power oscillation damping (POD).
As power systems are prone to topology changes because of connection and disconnection of generators and electrical lines, one of the most important requirements in the POD controller design is to account for these changes and to deal with them by using either adaptive or robust approaches. 
The robust approach is usually preferred by system operators because of the fixed structure of the controller.
In this paper, a procedure to design POD controllers for CIG-based power plants that takes into consideration all possible network configurations is presented. 
This procedure is based on frequency-response techniques, so it is suitable for the commissioning in newly installed power plants, even in those cases when a detailed small-signal model of the system is not available.
This procedure can be used to damp critical system modes by using active power, reactive power, or both power components simultaneously.
The proposed procedure is applied to the design of the POD controller for a CIG-based power plant connected to the IEEE~39 Bus system.
Simulations performed in Matlab and SimPowerSystems are used to validate the proposed design procedure.
\end{abstract}
\end{frontmatter}
\section{Introduction}
\label{Sec.introduction}
Low-frequency oscillations are inherent phenomena in power systems and are mainly caused by interactions between synchronous generators~(SG)~\cite{kundur1994power}.
To address these issues, the control system of SGs include a power system stabiliser~(PSS) loop~\cite{IEEE_PSS}.
The PSS acts upon changes in the grid frequency and adjusts the voltage reference of the SG.
This allows SGs to provide power oscillation damping~(POD) action.
However, the projected large-scale integration of converter-interfaced power plants introduces uncertainty with respect to POD in future power systems.
In their basic, conventional configuration, renewable energy plants do not provide POD services.
This change in the operation paradigm deteriorates the stability properties of power systems~\cite{Shah_2013_1}.
Furthermore, as the percentage of converter-interfaced power plants increases, this issue becomes more relevant~\cite{IEA_data}.
Thus, the trend is that transmission system operators now request from newly commissioned converter-interfaced power plants to participate in POD services~\cite{GridCodeREE},~\cite{New_Zeleand_Grid_Code}.

The initial proposal for providing POD services with converter-interfaced power plants was inspired by the operating principle of the PSS and it only used reactive power~\cite{Larsen1981}.
However, recent proposals show that both active and reactive powers can be used for that purpose~\cite{Liu2016},~\cite{Sun_2021}.
Furthermore, the combined action of both power components can further improve the damping action~\cite{Basu2021}, although negative interactions between two control loops may occur~\cite{Rimorov2015}.
Hence, wind power plants~\cite{Knuppel2012}, PV power plants~\cite{Sun_2021}, and battery systems~\cite{Yongli_2019} can all contribute to damping of critical system modes.
However, during the system operation, the changes in the system can lead to a shift in these modes.
These changes can be caused by loss of generation or demand elements, as well as by network reconfiguration.
Thus, the POD controller should take into consideration the changes caused by system modifications.

\begin{figure*}[!t]
\includegraphics[width=\textwidth]{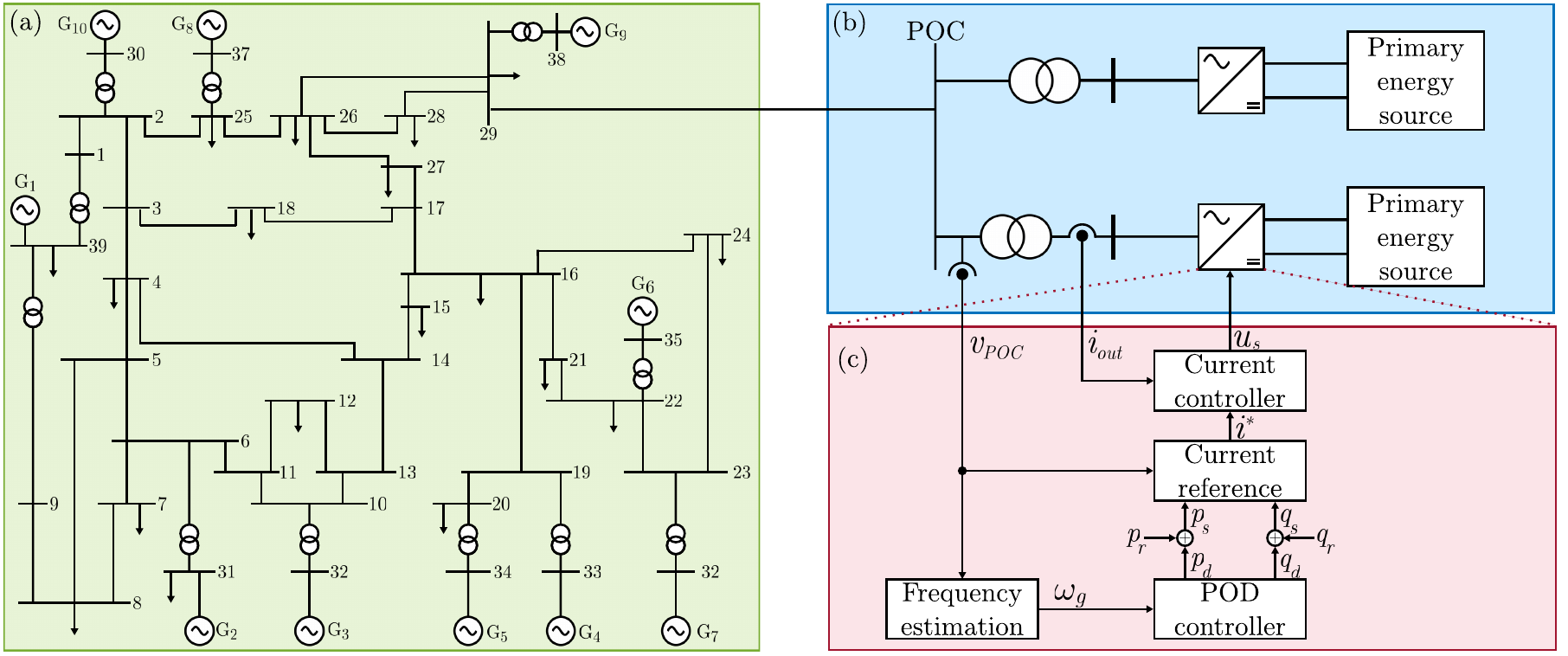}
\caption{Diagram of (a) IEEE 39 bus network, (b) Converter-interfaced power plant, and (c) control diagram for one converter.}
\label{Fig.system_overview}
\end{figure*}

One approach to address changes in the power system is to design an adaptive POD controller~\cite{Adjust_POD}.
Several proposals have been made with respect to this.
In most of them, different aspects of the power system operation are used for adjusting the POD controller.
For example, in~\cite{Chau_2018} the POD controller parameters are updated based on the load forecast.
Yao~\etal~\cite{Yao_2016} propose a predictive POD controller to address changes of the network operating point.
On the other hand, in~\cite{Arif_2014} uncertainty of the network operation is addressed by using a self-tuning POD controller that includes a neural network for estimating the power system behaviour.
Furthermore, the POD controller can adapt to changes in the network operating point~\cite{Deng_2019}.
Adaptive POD controllers are also suitable for addressing uncertainties arising from the power system model~\cite{Arif_2014}, and for uncertainties caused by the primary energy generation~(e.g., wind speed)~\cite{Zhang_2021}.
These proposals show that system stability margins can be improved by adjusting the POD controller parameters even under severe system variations.
However, the newly calculated parameter values may not be precise due to the discrepancy between the models used in the design and the real system.

Another approach to addressing uncertainties in the power system operation is to design a robust POD controller.
To do so, Konara~\etal~\cite{Konara_2016} propose a design based on eigenstructure assignment.
In this paper, the POD controller helps damp critical modes for different operating conditions.
In~\cite{Shah_2014}, a robust POD controller with a simple fixed second-order structure is presented, while in~\cite{Liu_2017} the POD controller structure is modified during the operation.
To improve the system robustness, the POD controller can be designed by solving an optimisation problem.
In~\cite{Yuan_2020} the objective function takes into consideration the transient stability aspects as well.
The same problem was also addressed by considering both regional pole placement and $\mathcal{H}_2$ performance metrics~\cite{Deng_2019}.
These proposals show that the POD controller can contribute to the damping of critical modes under variations in the system, such as changes in the system operating point.
Although these variations have an impact on dynamic properties, they do not include aspects such as a disconnection of system elements, despite  them having an important impact on both the operating point and the dynamic properties.

In this paper, a procedure for the design of a network-reconfiguration-aware POD controller for newly-commissioned power plants is proposed.
The procedure is based on the open-loop phase compensation method and relies on the system frequency response.
In this procedure, the frequency response of the power system for a large set of possible network configurations is analysed.
This set represents the realistic operating scenarios, in which one line at a time is disconnected.
Then, all this data is used to define an optimization problem to configure the parameters of a standard POD controller based on multiple series lead-lag filters.
After that, the system stability is examined for different values of controller gains.
Finally, the selected design is the one that achieves the best performance when applied to all network configurations under consideration.
The performance of the proposed procedure is tested and verified using the IEEE~39 Bus system benchmark network~\cite{IEEE_Benchmark}.
During the tests, different electrical lines were disconnected (one at the time) to demonstrate the system reconfiguration from the nominal case.
Results obtained using Matlab-SimPowerSystems~\cite{MATLAB_2019} are presented to validate the main findings of this work.
\begin{figure}[!b]
\centering
\includegraphics[width=\columnwidth]{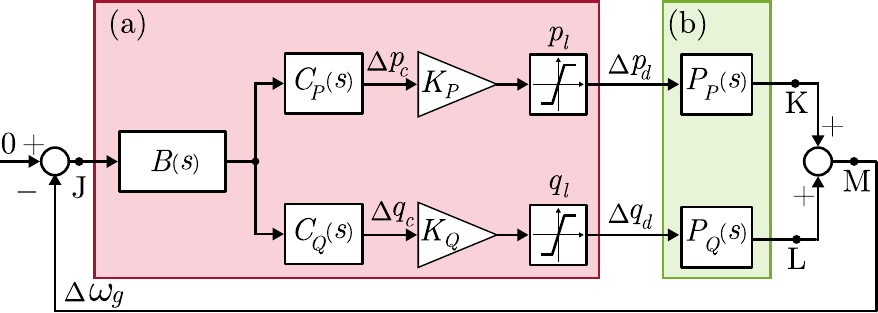}
\caption{Block diagram representation of (a) POD controllers and (b) plant models.}
\label{Fig.block_diagram}
\end{figure}

\begin{figure*}[!t]
\centering
\includegraphics[width=0.99\textwidth]{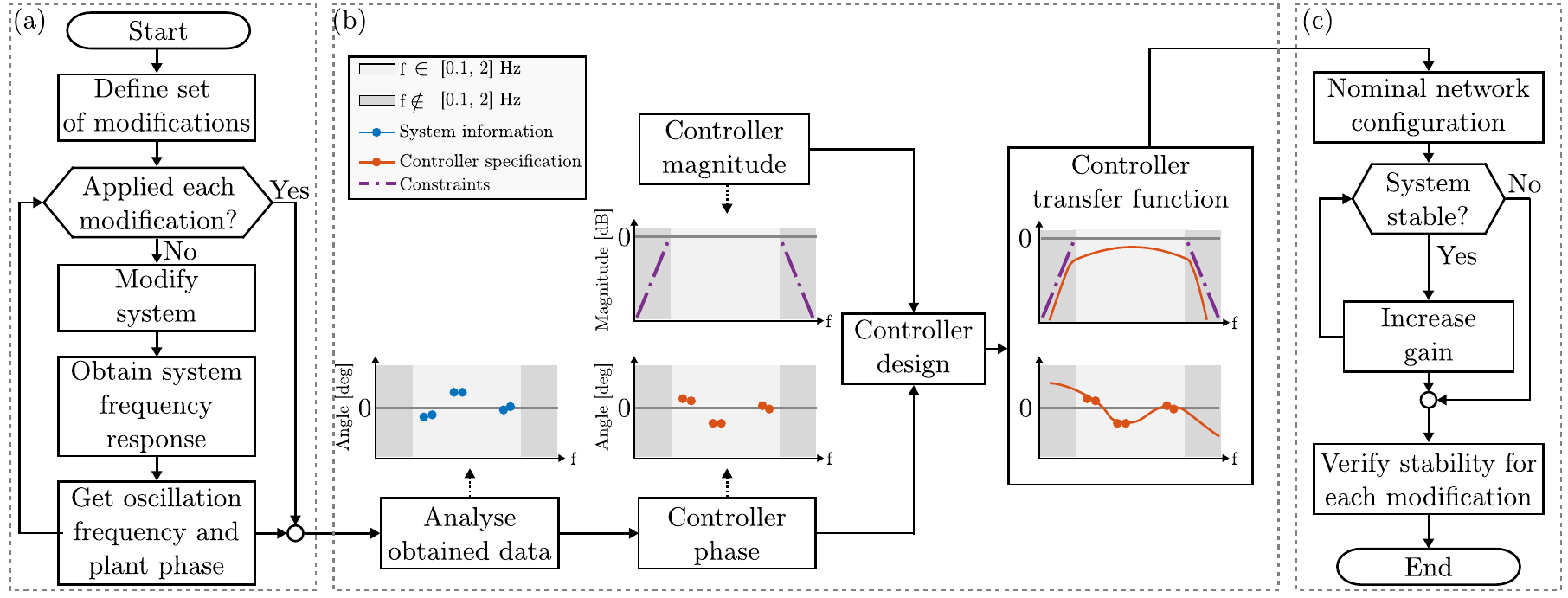}
\caption{Procedure for the design of network-reconfiguration-aware POD controller. 
(a) Network analysis, (b) design of compensators, and (c) proportional gain design.}
\vspace{-0.2cm}
\label{Fig.flowchart}
\end{figure*}

\section{System Overview and Methodology}
\label{Sec.system_overview}
\subsection{System Description}
Fig.~\ref{Fig.system_overview} shows the diagram of the power plant and the electrical system studied in this work.
It can be divided into three main parts.
The first part is depicted in Fig.~\ref{Fig.system_overview}~(a) and represents a model of a transmission network formed by ten generation units ($G_1$ to $G_{10}$).
Each generation unit includes a synchronous generator with a governor, an exciter and a PSS.
This model represents the IEEE~39~Bus network that is widely used as a benchmark network for inter-area oscillation studies~\cite{IEEE_Benchmark}.
In this work, two modifications are made to the original model.
First, several PSS are adjusted to reduce the damping of certain modes while keeping the system stable.
The second one is an additional point of connection~(POC) at bus 29 (see Fig.~\ref{Fig.system_overview}~(a)), which is used to connect a newly installed converter-interfaced power plant.

Fig.~\ref{Fig.system_overview}~(b) shows the electrical diagram of two additional converter interfaced generation units, representing the newly-commissioned power plant.
It consists of two converters used to connect the primary energy sources to the network.
These converters are connected to the POC via step-up transformers.
The output current~($i_{out}$) and voltage~($v_{POC}$) are measured and used in the control algorithm.

Fig.~\ref{Fig.system_overview}~(c) shows the control algorithm for one converter.
First, the grid frequency~($\omega_g$) is estimated from the measured voltage.
Then, the POD controller acts upon changes in the estimated frequency and produces references of active~($p_d$) and reactive~($q_d$) powers.
The total power references~($p_s$ and $q_s$) are calculated by adding the POD controller references and the references set by the system operator~($p_r$ and $q_r$).
The power references are then converted to the references of the current controller by taking into account the grid voltage at POC ($v_{POC}$). 
Finally, it is the current controller that defines the converter output voltage reference~($u_s$).
\subsection{POD Controller Overview}
Fig.~\ref{Fig.block_diagram} shows a block diagram of a small-signal representation of the system.
The POD controller block diagram is depicted in Fig.~\ref{Fig.block_diagram}~(a), and the controller plant models are depicted in Fig.~\ref{Fig.block_diagram}~(b).
In the plant, active and reactive powers are considered as inputs and the frequency is considered as an output.
The POD controller includes several elements.
The compensators $C_P(s)$ and $C_Q(s)$ consist of a series of lead-lag filters.
Besides those, band-pass filters, proportional gains and saturation blocks are added.
First, the band-pass filter is acting upon deviation of estimated network frequency $\Delta \omega_g$.
It removes components of $\Delta \omega_g$ outside the frequency range of interest.
Then, the lead-lag filter compensates the open-loop phase of the plant in order to maximise the damping action of the POD controller.
The outputs of these controllers are active~($p_c$) and reactive~($q_c$) power references.
The gains $K_P$ and $K_Q$ define the proportional action.
Finally, the limiting actions of saturation blocks in the active~($p_l$) and reactive~($q_l$) powers ensure the capabilities of the power plant are not exceeded.
This procedure produces the active~($p_d$) and reactive~($q_d$) power references for POD.
\subsection{POD Controller Design Methodology}
Fig.~\ref{Fig.flowchart} shows a flow diagram of the proposed procedure for designing network-reconfiguration-aware POD controllers.
This procedure is divided into three routines, described in the following sections.
\subsubsection{Network Analysis}
The network analysis is used to understand how the plants $P_P(s)$ and $P_Q(s)$ change when the network is reconfigured.
This routine is depicted in Fig.~\ref{Fig.flowchart}~(a).
First, the set of modifications considered during the design procedure is defined.
In this work, modifications include the original network topology with each of the electrical lines disconnected, one by one. 
This set of modifications can be extended to include cases such as loss of generation units or loads in the network.
Furthermore, the set of modifications can include changes in the network loading conditions or system operating point.
Such changes can emphasise nonlinear properties in the system.
Thus, these changes can be used to study the system's dynamic properties which might not be observed with changes in the network configuration.
These cases are also of interest, but they will not be addressed here for simplicity.
The next step in the plant analysis is to apply a series of network modifications and obtain the frequency response of $P_P(s)$ and $P_Q(s)$ for each configuration.
These transfer functions can be obtained from a detailed small-signal model.
However, if small-signal models are not available, $P_P(s)$ and $P_Q(s)$ can be obtained by using system identification techniques~\cite{Adjust_POD}.
Once these models are obtained, the peaks in the frequency response (i.e., resonance frequencies) are extracted.
The phases at the frequencies of the conflicting modes (i.e., phase to be compensated by POD controllers) are also calculated and stored.
This procedure is repeated for each modification in the given set of modifications.
\subsubsection{Compensator Design}
This routine includes the steps for designing the POD controller compensator, and it is depicted in Fig.~\ref{Fig.flowchart}~(b).
First, the information obtained in the previous routine is sorted and analysed.
In this step, the modifications that cause a significant change in the system dynamics~(e.g.,~loss of lines connected to the power plant) are neglected.
The remaining set of information is used to define the desired phase of the POD controller frequency response.
In addition, two constraints are added to shape the magnitude of the frequency response.
These constraints specify the maximum controller gain for the frequencies outside the frequency range of interest.
In any case, the controller impact on the system outside the specified frequency range is reduced.
This is of importance when the system frequency response is obtained by using the system identification approach.
In that case, the information about the system is determined by the characteristics of the perturbation signal.
The perturbation signal is selected to obtain information within the frequency range of interest.
Therefore, there is scarce information about the system dynamics outside the frequency range of interest.
Then, the controller design is defined as an optimisation problem in which the error between the specified and obtained phases is minimised.
The constraints for the optimisation problem are the aforementioned requirements for the magnitude.
\subsubsection{Design Verification}
The final routine for the design is depicted in Fig.~\ref{Fig.flowchart}~(c).
First, the system stability is examined for the nominal network configuration~(i.e., all lines connected) with the POD controller included.
Then, the controller gains are modified and the system response is obtained.
If the system encounters instability or any other practical reason that limits the controller gain (e.g., noise amplification, etc.), the iterative procedure is stopped.
Finally, the system stability is verified for all network configurations.
\section{Plant Analysis}
\label{Sec.plant_analysis}
During the system operation, the network topology changes every time a line in the system is disconnected.
Such changes have an important impact on the dynamic properties of the network.
In~\cite{Klein1991}, it has been shown that the change in the impedance of the lines affects the frequency of low-frequency modes.
In terms of frequency response, those changes have a two-fold effect~\cite{Adjust_POD}.
First, the changes in the frequency of critical modes are seen as changes in the oscillation frequency.
This means that the peak in the magnitude of the frequency response occurs at different frequencies.
The second effect is the change in the phase of the frequency response.

From the system definition in Fig.~\ref{Fig.block_diagram}, the frequency responses of $P_P(s)$ and $P_Q(s)$ can be defined as:
\begin{align}
P_P(j \omega_{o, i})
&=
A_{P, i} e^{j \phi_{P, i}},
\label{Eq.plant_p}\\
P_Q(j \omega_{o, i})
&=
A_{Q, i} e^{j \phi_{Q, i}},
~~\omega_{o, i} \in \Omega,
~~\textit{i} \in \mathbf{I},
\label{Eq.plant_q}
\end{align}
where $\omega_o$ represents the frequency of oscillations, while $A$ and $\phi$ represent the magnitude and phase of the frequency response at $\omega_o$, respectively.
The set $\mathbf{I}$ defines the number of oscillation frequencies.
From now, $i$ refers to an element in the set $\mathbf{I}$.
The set $\Omega$ represents oscillation frequencies of interest.
Then, the phases of the plants at these frequencies (expressions~(\ref{Eq.plant_p}) and~(\ref{Eq.plant_q})) are organised in corresponding sets as:
\begin{align}
\Phi_{P,n}
&=
[\phi_{P, 1},~\phi_{P, 2},~...,~\phi_{P, i}],\\
\Phi_{Q,n}
&=
[\phi_{Q, 1},~\phi_{Q, 2},~...,~\phi_{Q, i}].
\end{align}
Sets $\Phi_{P,n}$, $\Phi_{Q,n}$ are calculated for each network configuration~$\mathcal{N}=1,\;2,\;...,\;N$, where $N$ is the number of cases considered.
From now on, $n$ refers to an element in a given set of network configurations~$\mathcal{N}$.
Therefore, the set $\Omega_n$ should also bear the subscript $n$.
Next, the information from all modifications is organised and grouped as:
\begin{align}
\mathbf{\Omega}
&=
[\Omega_1,~\Omega_2,~...,~\Omega_n],\\
\mathbf{\Phi_{P}}
&=
[\Phi_{P, 1},~\Phi_{P, 2},~...,~\Phi_{P, n}],\\
\mathbf{\Phi_{Q}}
&=
[\Phi_{Q, 1},~\Phi_{Q, 2},~...,~\Phi_{Q, n}],
~~ \forall\textit{n} \in \mathcal{N},
\end{align}
where $\mathbf{\Omega}$, $\mathbf{\Phi_{P}}$, and $\mathbf{\Phi_{Q}}$ are sets that contain oscillation frequencies and phases for all the possible network configurations.
To design the POD controllers, the values in the sets $\mathbf{\Phi_{P}}$ and $\mathbf{\Phi_{Q}}$ will also include the phase shift introduced by additional filters added in the loop~(for more details, see Sec.~\ref{Sec.structure}).
\section{POD Controller Design}
\subsection{Controller Structure}
\label{Sec.structure}
A band-pass filter~($B(s)$) is applied to attenuate the signal content outside the range of low-frequency oscillations~--~$[0.1, 2]$~Hz.
It consists of high-pass and low-pass filters connected in series and it is defined as:
\begin{equation}
B(s)
=
\frac{s}{s + 1/T_h}
~
\frac{1}{s/T_l + 1},
\end{equation}
where $T_h$ and $T_l$ are time constants of high-pass and low-pass filters, respectively.
The values of $\omega_h$ and $\omega_l$ are selected taking into account the frequency range of low-frequency oscillations.
Consequently, these cut-off frequencies need to be selected so that $B(s)$ attenuates the signal frequency content outside the range of low-frequency oscillations.
At the same time, $B(s)$ will introduce an additional phase shift into the system, which needs to be added in sets $\mathbf{\Phi_P}$ and $\mathbf{\Phi_Q}$ so that it is compensated by $C_P(s)$ and $C_Q(s)$.

Compensators $C_P(s)$ and $C_Q(s)$ consist of a set of lead-lag filters connected in series.
$C_P(s)$ and $C_Q(s)$ have the same structure and their parameters are defined following the same procedure.
Thus, in the rest of the document, the developments are presented only for $C_P(s)$ (for simplicity).
The controller $C_P(s)$ is defined as:
\begin{equation}
C_P(s)
=
\frac{1+sT_1}{1+sT_2}
\cdot
\frac{1+sT_3}{1+sT_4}
\cdot
...
\cdot
\frac{1+sT_{c-1}}{1+sT_{c}},
\label{Eq.tf_controller}
\end{equation}
where $T_{1}-T_{c}$ are the compensator time constants, while $c$ represents the number of time constants.

The order of the compensator~(\ref{Eq.tf_controller}) should be selected, and it depends on two requirements.
First, the requirements for the phase of $C_P(j\omega)$ are defined to compensate for a set of phases obtained in the plant analysis.
This set of phases includes a number of points with large phase differences across a narrow frequency range.
Therefore, a low-order transfer function may lead to a large deviation between the required phase and the phase introduced by $C_P(j \omega)$.
On the other hand, the results achieved with very high-order transfer functions would not necessarily result in significantly improved design performance.
At the same time, higher-order transfer functions are typically difficult to implement in real-time~\cite{Lyons2004}.
Thus, the order $C_P(s)$ represents a compromise between these two considerations.
\subsection{Design Objective}
The method used to design the POD controllers is based on the phase compensation of the open-loop plant at the frequency of interest~\cite{Ogata2010modern}.
The open-loop transfer function~($G(s)$) is defined as a transfer function from point J to point M, in Fig.~\ref{Fig.block_diagram}.
Analytically, the compensation objective can be written as:
\begin{equation}
\phi_G = 0,
\label{Eq.open_phase}
\end{equation}
\noindent where
\begin{equation}
G(j\omega_o)
= A_G e^{j \phi_G}.
\end{equation}
It has already been shown that this design method provides a system damping that is close to the optimal value~\cite{ISGT_POD}.
Also, as the design is based on open-loop characteristics, the calculation of the compensator parameters can be easily automatised.

Two open-loop transfer functions can be calculated from  Fig.~\ref{Fig.block_diagram}.
These transfer functions represent the active~($G_P(s)$) and reactive~($G_Q(s)$) power dynamics, and these are defined from point J to points K and M in Fig.~\ref{Fig.block_diagram}, respectively.
In~\cite{ISGT_POD}, it has been shown that the criteria from~(\ref{Eq.open_phase}) can be met by setting phases to zero for two open-loop functions separately:
\begin{equation}
\phi_{GP} = \phi_{GQ} = 0,
\label{Eq.two_open_phases}
\end{equation}
where $\phi_{GP}$ and $\phi_{GQ}$ are phases of $G_P(j \omega_o)$ and $G_Q(j \omega_o)$, respectively.

To meet the objectives presented in~(\ref{Eq.two_open_phases}), each controller ($C_P(s)$ and $C_Q(s)$) should compensate the phase of its corresponding open-loop system at the oscillation frequency~\cite{ISGT_POD}.
Taking into account multiple oscillation frequencies and changes caused by network reconfiguration, this criteria can be defined as (only the active power loop is shown):
\begin{equation}
\phi_{GP}^{i,n} 
=
\phi_{CP}^{i,n} - \phi_P^{i,n}
=
0,\;\;\;\;\;\;\forall i\in\mathbf{I},\;\forall n\in\mathcal{N},
\label{Eq.open_active}
\end{equation}
where $\phi_{CP}^{i,n}$ is phase of $C_P(j \omega)$.

The criteria from~(\ref{Eq.open_active}) means that the controller $C_P(s)$ needs to compensate the phase of the corresponding plant for all oscillation frequencies and all network configurations.
Such requirements are too strict, mainly for two reasons.
First, the requirements for the phase of $C_P(j \omega)$ would include many points in a very narrow frequency range~(0.1 to 2 Hz).
Second, plants for different network configurations might have the same frequency of oscillation with different phases.
Therefore, trying to fully compensate for the phase in all the cases is not a reasonable solution.
Instead, an alternative function to design the compensator is proposed in the following section.
\subsection{Optimisation Problem Definition}
A cost function is defined based on the criteria presented in~(\ref{Eq.open_active}):
\begin{equation}
\min_{{\mathbf{x}}}
\sum_{\substack{\textit{i} \in \mathbf{I}, \\ \textit{n} \in \mathcal{N}}}
\Big| \phi_{CP}^{i, n} - \phi_{P}^{i, n} \Big|^m,
\label{Eq.min}
\end{equation}
where
\begin{equation}
{\mathbf{x}}
=
[T_1, T_2, ..., T_{c}].
\end{equation}
The parameter $m$ defines the error cost, and its value depends on the complexity of the optimisation problem.
Namely, the change among elements~$\phi_P^{i,n} \in \mathbf{\Phi_P}$ can be different depending on the results of the plant analysis.
This change has a direct impact on the complexity of the problem in~(\ref{Eq.min}).
Therefore, the parameter $m$ is case specific, and its value depends on two criteria.
The first criteria is the allowed average error for the controller design.
This error determines the performance of the designed controller in damping multiple modes in various network configurations.
The second criteria is the maximum allowed error for any case.
Thus, the selection of the parameter $m$ value is a compromise between these two criteria.
\subsection{Constraints}
The solution of problem~(\ref{Eq.min}) defines the controller parameters based on the required phases in the frequency range of low-frequency oscillations.
Nonetheless, once the controller is introduced into the system, it affects the rest of the frequencies outside this range.
If not addressed, it may lead to unwanted interactions with other control layers.
To address this issue, constraints are added to the  optimisation problem.
These constraints ensure that the controller impact outside the range of low-frequency oscillations is lower than within this range and are defined as:
\begin{equation}
c
:
\big| C_P(j \omega_{out}) \big| \leq 1,
~~~~~~~\forall~\omega_{out} \in \mathbf{\Omega_{out}},
\label{Eq.constraints}
\end{equation}
where $\mathbf{\Omega_{out}}$ is a set of frequencies defined as:
\begin{equation}
\omega_{out}
\notin
\big[0.1, 2\big]\cdot2\pi~\textit{rad}/\textit{s},
~~\forall~\omega_{out} \in \mathbf{\Omega_{out}}.
\end{equation}
The constraints in~(\ref{Eq.constraints}) limit the magnitude of $C_P(j\omega)$.
It is relevant to mention the POD controller gain (i.e., the gain in the range of interest) will be tuned in the last step of the design procedure (with $K_P$). 

Another set of constraints is linked with the set of unknowns $\mathbf{x}$ in the optimisation problem~(\ref{Eq.min}).
Namely, the unknowns in this problem are controller parameters.
Thus, constraints for those parameters are necessary to guarantee that the obtained transfer function $C_P(s)$ is stable.
These constraints are defined as:
\begin{equation}
T_{k} > 0,
~~\forall~~\textit{k} \in \Big\{ 2, 4, 6, ...\Big\}.
\end{equation}
\subsection{Controller Gain}
\begin{figure}[!t]
\centering
\includegraphics[width=\columnwidth]{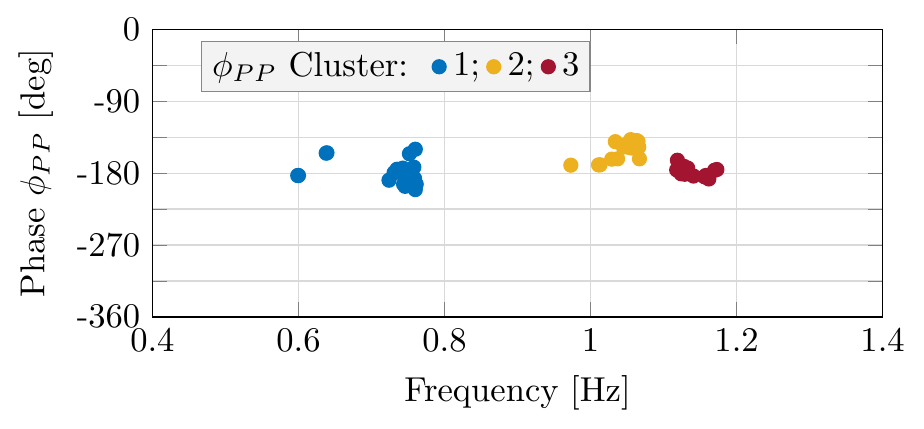}
\vspace{-0.6cm}
\caption{Phase of $P_P(j\omega)$ for different network configurations. (blue, yellow, red) Clusters of points from three oscillation frequencies.}
\label{Fig.plant_P}
\end{figure}
\begin{figure}[!t]
\centering
\includegraphics[width=\columnwidth]{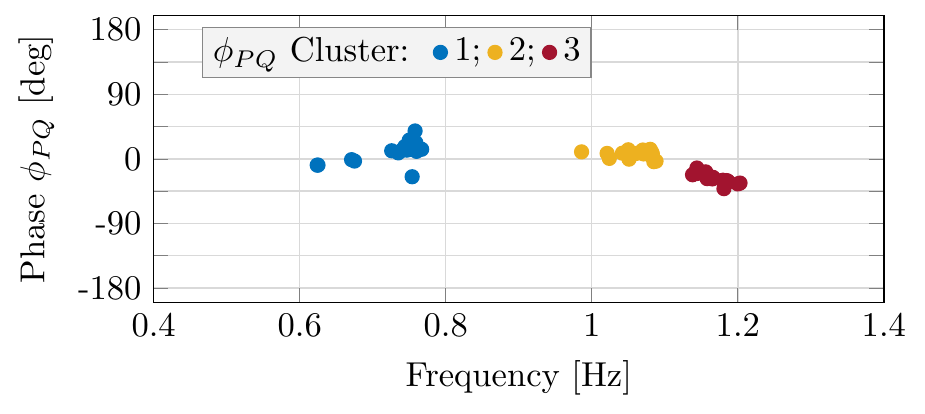}
\vspace{-0.6cm}
\caption{Phase of $P_Q(j\omega)$ for different network configurations. (blue, yellow, red) Clusters of points from three oscillation frequencies.}
\vspace{-0.4cm}
\label{Fig.plant_Q}
\end{figure}
\begin{table}[!b]
\centering
\vspace{-0.6cm}
\caption{Design errors $E_P$ and $E_Q$ for different number of lead-lag filters}
\include{JOURNAL/TABLES/Table_order_vs_error}
\label{Tab.order_vs_error}
\end{table}
The controller gains~($K_P$ and $K_Q$) are determined based on the system frequency response, following the procedure from Fig.~\ref{Fig.flowchart}~(c).
This procedure consists of two parts.
In the first part, the system is analysed for the case when all the lines are connected.
Gain values are changed incrementally and the system's closed-loop frequency response is observed.
The information from this analysis describes the impact of the gains on the system stability margins.
Such change results in improved damping of low-frequency modes.
Nonetheless, after a certain point, the damping of one or more modes would decrease.
Furthermore, as the gain increases, the impact of the POD controller outside the frequency range of interest is changing.
The change can be seen as a decrease in stability margins below and above the frequency range, depending on the system characteristics.
Therefore, the selection of the gain value takes into account different aspects of the controller impact on the system stability.

The second part of the procedure includes the verification of system stability with the gain value selected in the first step.
Here, the system stability is verified for all the network configurations using the selected gain value.
For this purpose, the set of plants for different network configurations can be used.
Since these plants are obtained in plant analysis~(Section~\ref{Sec.plant_analysis}), no additional identification procedure is required.
If the system is unstable for any network configuration, the first step of the procedure for the gain value selection needs to be repeated and a different gain value should be selected.
\subsection{Controller Limits}
During transients, the converter is delivering active and reactive powers for oscillation damping.
Depending on the amplitude of the frequency deviation, the power references from the POD controller can exceed the converter capacity.
It is, therefore, necessary to define the active and reactive power limits~($p_l$ and~$q_l$).
These limits are changing over time based on the network operating point and available energy from the primary energy source.
For simplicity, these limits are kept constant in this work (see~\cite{ISGT_POD} for more details).
\section{Results}
\subsection{Plant Analysis}
Fig.~\ref{Fig.plant_P} and Fig.~\ref{Fig.plant_Q} show the changes of elements in $\mathbf{\Phi_P}$ and $\mathbf{\Phi_Q}$ with respect to elements in $\mathbf{\Omega}$ for different network configurations.
In both cases, three clusters of points are formed, corresponding to three different ranges of oscillation frequencies.
These clusters are marked in blue, yellow and red in 
Fig.~\ref{Fig.plant_P} and Fig.~\ref{Fig.plant_Q}.
Changes in each cluster describe how network reconfiguration affects the interactions between different generators in the system.
It can be seen that the angle and frequency variations differ between the clusters.
Angles  $\phi_P$ and $\phi_Q$ vary around~90 degrees, while the frequency of oscillations varies in the range of~0.1~Hz.
\begin{figure}[!b]
\centering
\includegraphics[width=\columnwidth]{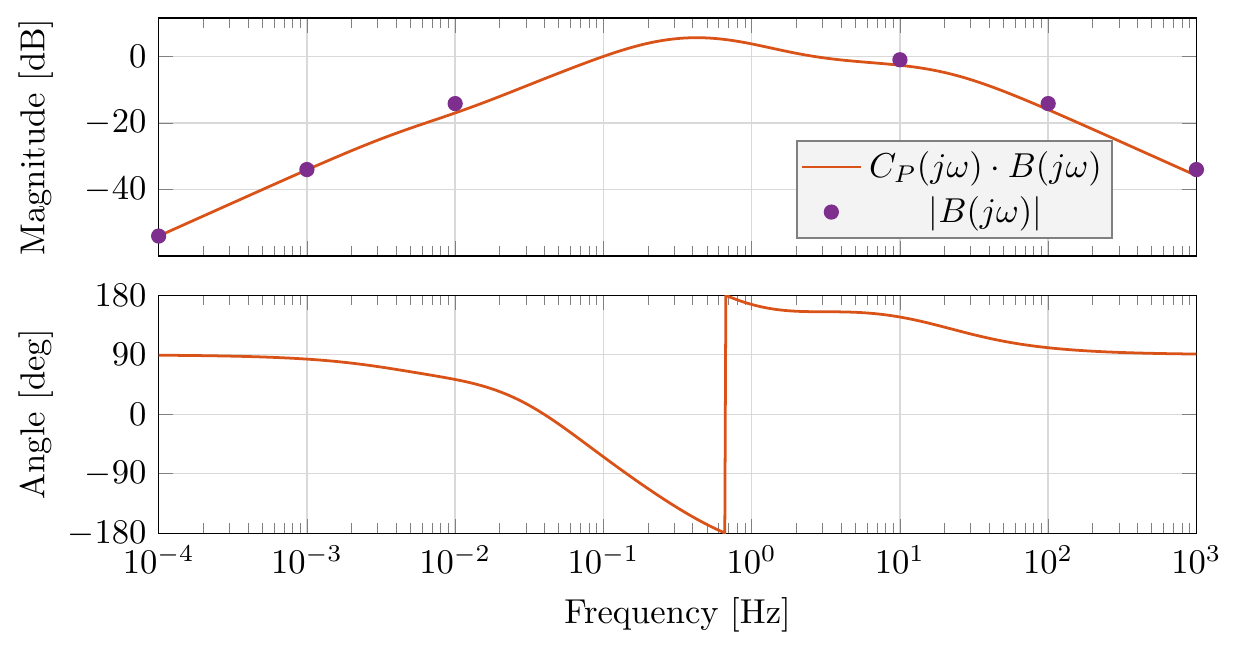}

\caption{(orange) Frequency response of ($C_P(s) \cdot B(s)$) and (purple) magnitude of $B(j \omega)$ for all elements in $\mathbf{\Omega_{out}}$.}
\label{Fig.CP_bode}
\end{figure}
\begin{figure}[!t]
\centering
\includegraphics[width=\columnwidth]{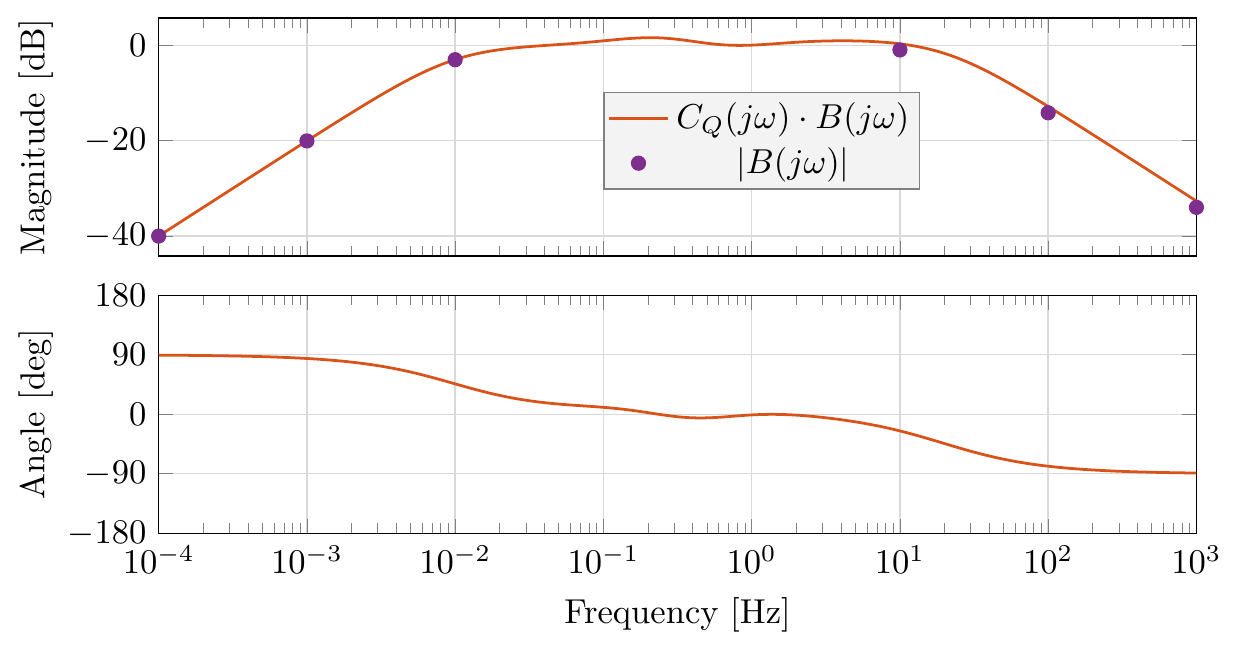}
\vspace{-0.9cm}
\caption{(orange) Frequency response of ($C_Q(s) \cdot B(s)$) and (purple) magnitude of $B(j \omega)$ for all elements in $\mathbf{\Omega_{out}}$.}
\vspace{-0.3cm}
\label{Fig.CQ_bode}
\end{figure}
\subsection{Compensator Design}
The compensators are designed using the proposed method using the information obtained from the plant analysis.
The optimisation problem~(\ref{Eq.min}) with constraints from~(\ref{Eq.constraints}) is solved using \textit{fmincon} function from Matlab~\cite{MATLAB_2019}.
The computational time required to design compensators $C_P(s)$ and $C_Q(s)$ was~4.5 and~3.4 minutes, respectively.
The optimisation problem is solved on desktop~PC with \textit{i5-8500} CPU with~8~GB RAM.

The optimisation problem~(\ref{Eq.min}) is solved using the parameter setting $m=3$.
The compensator design accuracy is verified using the error between the specified and the obtained phase of $C_P(j\omega)$ for all the elements in $\mathbf{\Omega}$.
This error is defined as:
\begin{equation}
E_P^{i,n}
=
\Big|\phi_{CP}^{i,n} - \phi_{P}^{i,n}\Big|
,\;\;\;\;\;\;\forall i\in\mathbf{I},\;\forall n\in\mathcal{N}.
\end{equation}
This error corresponds to the individual elements of the sum that is used to define the cost function~(\ref{Eq.min}).
Thus, by minimising the cost function, error $E_P^{i,n}$ is reduced.
\begin{figure}[!t]
\centering
\includegraphics[width=\columnwidth]{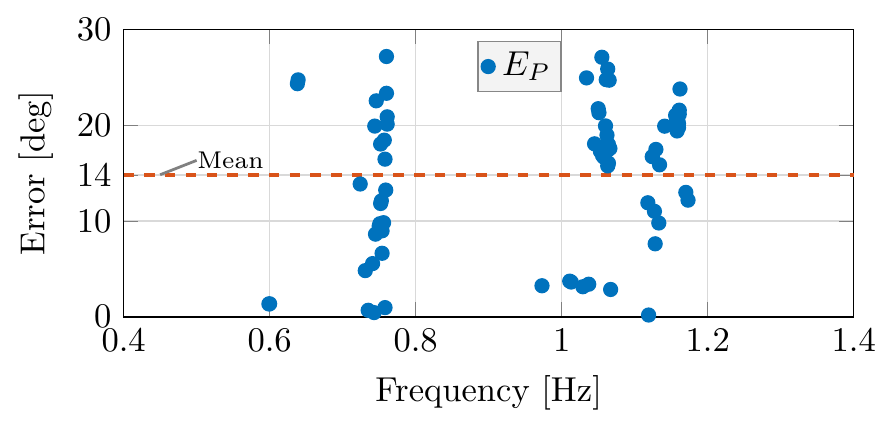}
\vspace{-0.8cm}
\caption{(blue) Error between the specified and achieved phase of $C_P(j\omega)$ and (orange) mean absolute error.}
\vspace{-0.3cm}
\label{Fig.CP_error}
\end{figure}
\begin{figure}[!t]
\centering
\includegraphics[width=\columnwidth]{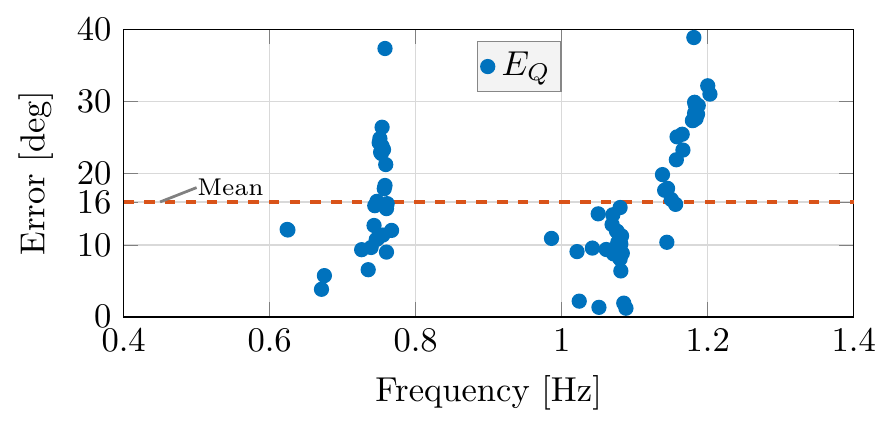}
\vspace{-0.8cm}
\caption{(blue) Error between the specified and achieved phase of $C_Q(j\omega)$ and (orange) mean absolute error.}
\vspace{-0.2cm}
\label{Fig.CQ_error}
\end{figure}

Table~\ref{Tab.order_vs_error} shows the mean and maximum errors for both compensators $C_P(s)$ and $C_Q(s)$ when different compensator orders are used.
For $C_P(s)$ it can be seen that the error is significantly reduced if a series of two lead-lag filters are used compared to the case with only one such filter.
Further increase in the number of lead-lag filters leads to a reduced design error, although the difference is not significant.
Then, for $C_P(s)$ of order five, the maximum error increases significantly, while the compensator of order six gives the same results as those of orders three and four.
On the other hand, the error in $C_Q(s)$ does not change significantly when the order increases.
By increasing the order from one to three, the design error is not modified.
Then, for orders four to six, the error is changed, although the improvement is not significant.
These results show that the relation between the compensator order and the design error is not linear.
Therefore, the order of the compensator transfer function should be selected carefully upon determining the relation between the error and the compensator order.
Based on the analysis, the orders for $C_P(s)$ and $C_Q(s)$ are set to 3 and 4, respectively.

Table~\ref{Tab.compensators} shows the time constants of lead-lag filters used in $C_P(s)$ and $C_Q(s)$, for the selected configuration.
For $C_Q(s)$ there are three lead-lag filters with similar parameters, although the optimisation problem did not have such specifications.
However, this result can be expected and it is similar to the common practice in designing PSS for synchronous generators.
\begin{figure}[!t]
\centering
\includegraphics[width=\columnwidth]{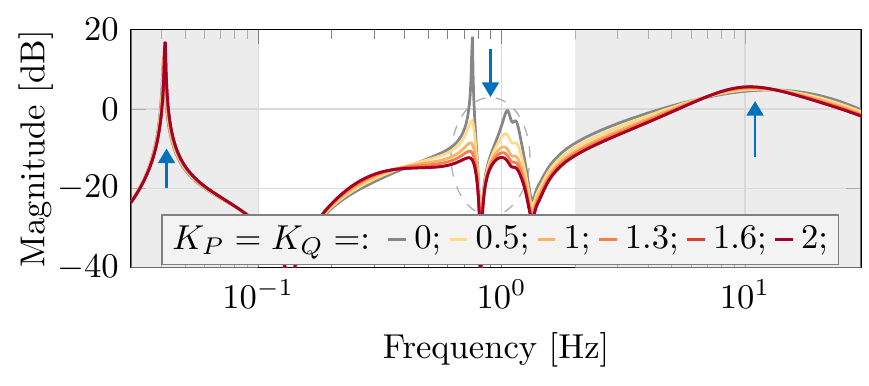}
\vspace{-0.9cm}
\caption{System closed-loop frequency response for variation of proportional gain values.}
\label{Fig.gain_bode}
\end{figure}
\begin{figure}[!t]
\centering
\includegraphics[width=\columnwidth]{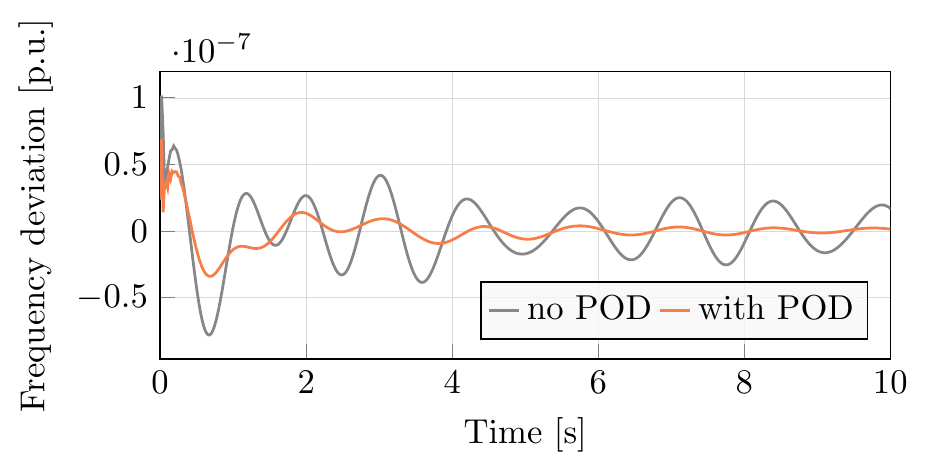}
\vspace{-0.8cm}
\caption{Transient response after a disturbance is applied (gray) without and (orange) with POD controller.}
\label{Fig.gain}
\end{figure}

Fig.~\ref{Fig.CP_bode} and Fig.~\ref{Fig.CQ_bode} show the frequency response of POD controllers designed using the proposed method.
These figures show the combined frequency response of the band-pass filter and compensators in each loop allowing to assess the total impact of the POD controller on the system.
It can be seen that the band-pass filter mainly determines the shape of the magnitude of the POD controller, due to the magnitude constraints of $C_P(j\omega)$ and $C_Q(j\omega)$.
Within the frequency range of interest, the POD controller magnitude deviates from the band-pass filter.
This is expected because in this frequency range the requirements for the phase of $C_P(j\omega)$ and $C_Q(j\omega)$ are specified and set.
\begin{table}[!t]
\centering
\small
\caption{Parameters for compensators $C_P(s)$ and $C_Q(s)$}
\include{JOURNAL/TABLES/Table_compensators}
\vspace{-0.4cm}
\label{Tab.compensators}
\end{table}
\begin{table}[!t]
\centering
\small
\caption{POD Controller Parameters}
\include{JOURNAL/TABLES/Table_parameters}
\vspace{-0.4cm}
\label{Tab.parameters}
\end{table}

Fig.~\ref{Fig.CP_error} and Fig.~\ref{Fig.CQ_error} show the design errors for both $C_P(s)$ and $C_Q(s)$, respectively.
Overall, it can be seen that the errors in both power loops are lower than 40~degrees, while in the majority of cases, it is even lower.
The mean error in active and reactive power loops is shown as an orange line in Fig.~\ref{Fig.CP_error} and Fig.~\ref{Fig.CQ_error}, respectively.
It can be seen that the mean errors are around 15~degrees in both cases.
This ensures the system is well-damped in most of the cases considered.
\subsection{Controller Gain}
Fig.~\ref{Fig.gain_bode} shows the system frequency response for modified values of gains $K_P$ and $K_Q$.
Here, the system is in nominal configuration, with all its lines connected.
It can be seen that the increase of the proportional gain has a twofold effect on the system's dynamic properties.
First, for three low-frequency peaks, the system magnitude decreases as the gain value increases.
At the same time, the change in the opposite direction is seen for peaks outside the frequency range of interest~(depicted as gray areas in Fig.~\ref{Fig.gain_bode}).
Thus, the system stability margins for frequencies outside the range of low-frequency oscillation, are reduced with an increase of the controller gains.
These results show the importance of analysing stability margins for frequencies within and outside the frequency range of interest.
Based on this analysis, the proportional gains are set to $K_P=K_Q=1.3$.
These values, together with the rest of the relevant system parameters are summarised in Table~\ref{Tab.parameters}.
Fig.~\ref{Fig.gain} shows the system transient response after a disturbance is applied without (gray) and with (orange) the POD controller.
A significant improvement in damping can be seen as the system reaches a steady state within a few seconds after disturbance, compared to tens of seconds for the system without a POD controller.
\subsection{System performance}
\begin{figure}[!t]
\centering
\includegraphics[width=\columnwidth]{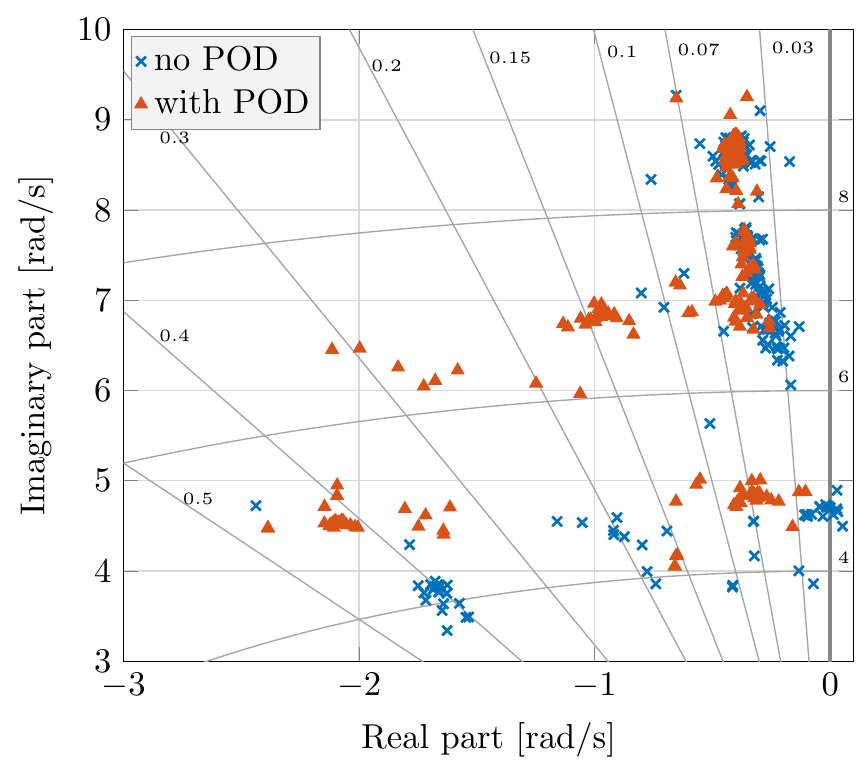}
\vspace{-0.7cm}
\caption{System eigenvalues for network configurations with different lines disconnected (cross) without POD controller and (triangle) with POD controller.}
\vspace{-0.4cm}
\label{Fig.system_eigs}
\end{figure}
Fig.~\ref{Fig.system_eigs} shows the relevant system eigenvalues for different network configurations (blue) without and (orange) with the POD controller.
Here, the system is observed for all possible different disconnections of lines in the network, including those that are excluded during the design procedure.
It can be seen that the system under the network reconfiguration without a POD controller faces significant changes of its stability margins.
For certain network topologies, the system even becomes unstable.
Once the POD controller is introduced, the change of the system eigenvalues is reduced and the damping factors are improved, for all possible network configurations.
Furthermore, the system remains stable once the network configuration is changed.
\vspace{-0.2cm}
\section{Conclusion}
\label{Sec.conclusions}
A method for design of a network-configuration-aware POD controller for newly commissioned power plants is presented in this paper.
The proposed design procedure takes into account the changes in the system caused by network reconfiguration.
In this paper it has been explained how all the necessary information for the proposed method is obtained from the system frequency response.
An optimisation based procedure for the POD controller parameter design has been introduced.
The formulation of the optimisation problem for this purpose including all the system constraints has been proposed and its objective defined.
All the steps of the proposed design procedure and its validation have been explained in detail. 
The proposed design method was tested using the IEEE~39 Bus system in Matlab-SimPowerSystems~\cite{MATLAB_2019} by disconnecting all the network lines (one at the time).

The obtained results have confirmed that the changes in the network configuration have an important effect on the dynamic properties of the system and that they need to be accounted for during the POD controller design procedure.
While in majority of the cases the changes resulted in a reduction of system stability margins, in some cases the power system was unstable.
Once the network-reconfiguration-aware POD controller was introduced, the changes in the network topology had limited impact on the system stability margins.
It has been shown that the robustness of the controller increases with the number of lead-lag filter stages, although the controller performance barely increases after a certain number of stages.
Also, the results have demonstrated that the system eigenvalues remain well-damped for all the possible cases studied, highlighting the principal advantage of the proposed POD controller design.

Future work will focus on extending the set of considered cases with modifications caused by the loss of generation units and variations in the network loading conditions.
Also, it is of interest to study the performance of the proposed method applied to different networks because the relation between the oscillation frequency and the compensation angles depends on the network topology.
\vspace{-0.2cm}
\bibliography{cas-refs}
\bibliographystyle{elsarticle-num}
\end{document}

%% file: JOURNAL/FIGURES/Style.tex
\usepackage{tikz}
\usetikzlibrary{positioning,shapes,arrows.meta,patterns,calc}
\usepackage[european,american inductors,siunitx]{circuitikz}
\usepackage{pgfplots}
\usepgfplotslibrary{groupplots}
\usetikzlibrary{spy}
\pgfdeclarelayer{points}
\pgfsetlayers{main,points}
\pgfplotsset{compat=1.15}

\definecolor{Tred}{rgb}{0.68, 0.05, 0.0}
\definecolor{Tgreen}{rgb}{0.0, 0.5, 0.0}
\definecolor{Torange}{rgb}{0.95, 0.52, 0.04}
\definecolor{Tyellow}{rgb}{1.0, 0.73, 0.0}
\definecolor{Tblue}{rgb}{0.28, 0.57, 0.81}

\definecolor{blue_m}{rgb}{0, 0.4470, 0.7410}
\definecolor{orange_m}{rgb}{0.8500, 0.3250, 0.0980}
\definecolor{yellow_m}{rgb}{0.9290, 0.6940, 0.1250}
\definecolor{purple_m}{rgb}{0.4940, 0.1840, 0.5560}
\definecolor{green_m}{rgb}{0.4660, 0.6740, 0.1880}
\definecolor{lightblue_m}{rgb}{0.3010, 0.7450, 0.9330}
\definecolor{blue_m}{rgb}{0, 0.4470, 0.7410}
\definecolor{red_m}{rgb}{0.6350, 0.0780, 0.1840}
\definecolor{gray_m}{rgb}{0.529, 0.529, 0.529}
\definecolor{gray_m_fill}{rgb}{0.952, 0.952, 0.952}

\definecolor{red4}{rgb}{0.698, 0.094, 0.169}
\definecolor{red3}{rgb}{0.839, 0.376, 0.302}
\definecolor{red2}{rgb}{0.957, 0.647, 0.51}
\definecolor{red1}{rgb}{0.992, 0.859, 0.78}
\definecolor{white}{rgb}{0.969, 0.969, 0.969}
\definecolor{blue1}{rgb}{0.820, 0.890, 0.940}
\definecolor{blue2}{rgb}{0.573, 0.773, 0.871}
\definecolor{blue3}{rgb}{0.263, 0.576, 0.765}
\definecolor{blue4}{rgb}{0.129, 0.4, 0.675}

\definecolor{orange_q}{rgb}{0.933, 0.466, 0.2}
\definecolor{blue_q}{rgb}{0, 0.466, 0.733}
\definecolor{cyan_q}{rgb}{0.2, 0.733, 0.933}
\definecolor{magenta_q}{rgb}{0.933, 0.2, 0.466}
\definecolor{red_q}{rgb}{0.8, 0.2, 0.066}
\definecolor{teal_q}{rgb}{0, 0.6, 0.533}
\definecolor{gray_q}{rgb}{0.733, 0.733, 0.733}

\definecolor{sunset_1}{HTML}{364B9A}
\definecolor{sunset_2}{HTML}{4A7BB7}
\definecolor{sunset_3}{HTML}{6EA6CD}
\definecolor{sunset_4}{HTML}{98CAE1}
\definecolor{sunset_5}{HTML}{C2E4EF}
\definecolor{sunset_6}{HTML}{EAECCC}
\definecolor{sunset_7}{HTML}{FEDA8B}
\definecolor{sunset_8}{HTML}{FDB366}
\definecolor{sunset_9}{HTML}{F67E4B}
\definecolor{sunset_10}{HTML}{DD3D2D}
\definecolor{sunset_11}{HTML}{A50026}

\definecolor{step_1}{HTML}{125A56}
\definecolor{step_2}{HTML}{00767B}
\definecolor{step_3}{HTML}{238F9D}
\definecolor{step_4}{HTML}{42A7C6}
\definecolor{step_5}{HTML}{60BCE9}
\definecolor{step_6}{HTML}{9DCCEF}
\definecolor{step_7}{HTML}{C6DBED}
\definecolor{step_8}{HTML}{DEE6E7}
\definecolor{step_9}{HTML}{ECEADA}
\definecolor{step_10}{HTML}{F0E6B2}
\definecolor{step_11}{HTML}{F9D576}
\definecolor{step_12}{HTML}{FFB954}
\definecolor{step_13}{HTML}{FD9A44}
\definecolor{step_14}{HTML}{F57634}
\definecolor{step_15}{HTML}{E94C1F}
\definecolor{step_16}{HTML}{D11807}
\definecolor{step_17}{HTML}{A01813}

\tikzset{
    bus/.style =  {
                    draw,
                    line width=0.3mm,
                    line cap=round},
    load/.style = {
                    draw,
                    thick,
                    regular polygon,
                    regular polygon sides=3,
                    shape border rotate=180,
                    minimum width=1mm,
                    inner sep=1.5pt,
                    fill=white},
    grid_line/.style = {
                        draw,
                        line cap=round},
    PE_block/.style =  {
                        draw,
                        rectangle,
                        anchor=east,
                        minimum height = 1cm,
                        minimum width = 1cm},
    block/.style =  {
                    draw,
                    rectangle,
                    minimum height = 2em,
                    minimum width = 2em,
                    inner sep=0.4mm,
                    outer sep=0mm,},
    sum_PN_1/.style =   {
                        draw,
                        circle,
                        minimum width = 0.4cm},
    saturation/.style={
                        draw,
                        minimum height = 2em,
                        minimum width = 2em,
                        path picture={
                                        \pgfpointdiff{
                                            \pgfpointanchor{path picture bounding box}{north east}}
                                        {\pgfpointanchor{path picture bounding box}{south west}}
                                        \pgfgetlastxy\x\y
                                        \tikzset{x=\x*.4, y=\y*.4};
                                        \draw [ultra thin] (-1,0) -- (1,0) (0,-1) -- (0,1); 
                                        \draw (-1,-.7) -- (-.4,-.7) -- (.4,.7) --      (1,.7);}},
    droop/.style={
                        draw,
                        minimum height = 2em,
                        minimum width = 2em,
                        path picture={
                                        \pgfpointdiff{
                                            \pgfpointanchor{path picture bounding box}{north east}}
                                        {\pgfpointanchor{path picture bounding box}{south west}}
                                        \pgfgetlastxy\x\y
                                        \tikzset{x=\x*.4, y=\y*.4};
                                        \draw [ultra thin] (0.7,0.7) -- (-0.7,0.7) (0.7,0.7) -- (0.7,-0.7);
                                        \draw (0.958,-0.4) -- (-0.7,0);
                                        }},
    gain/.style =  {
                    draw,
                    isosceles triangle,
                    isosceles triangle stretches,
                    anchor=left corner,
                    inner sep=0.5mm,
                    outer sep=0mm,},
    quest/.style = {
                    diamond,
                    draw,
                    minimum height = 1cm,
                    minimum width = 1cm,
                    text centered, 
                    aspect=2},  
    -|-/.style={
                to path={
                        (\tikztostart) -|($(\tikztostart)!#1!(\tikztotarget)$) |-(\tikztotarget)
                        \tikztonodes}},
    -|-/.default=0.5,
    set point size/.code={\pgfmathsetmacro{\pointsize}{sqrt(\pgflinewidth)}},
    point size/.style={set point size/.prefix style={line width=#1}},
    every dot/.style = {inner sep=0, outer sep=0,font=},
    outer dot/.style = {every dot, scale=4*\pointsize},
    inner dot/.style = {every dot, scale=\pointsize, text=#1}, inner dot/.default={black},
    point fill/.style = {inner dot/.default={#1}},
    point coordinate/.style={insert path={coordinate(#1)}},
    point/.style={insert path={
      node[inner sep=0, overlay, every point/.try, #1, set point size,
        path picture={
          \begin{pgfonlayer}{points}
            \node[outer dot]{.} node[inner dot]{.};
          \end{pgfonlayer}
        }
      ]
    }
  }
}

%% file: JOURNAL/TABLES/Table_order_vs_error.tex
\begin{tabular}{@{}ccccc@{}}
\toprule
\multicolumn{1}{c}{Order}
& \multicolumn{2}{c}{$E_P$}
& \multicolumn{2}{c}{$E_Q$}
\\
\cmidrule(l){2-3}
\cmidrule(l){4-5}
& mean
& max
& mean
& max
\\
1
& 164
& 179
& 17
& 41
\\
2
& 15
& 27
& 17
& 41
\\
\cellcolor[HTML]{e8f4d7}3
&\cellcolor[HTML]{e8f4d7}14
&\cellcolor[HTML]{e8f4d7}27
& 17
& 41\\
\cellcolor[HTML]{e8f4d7}4
& 14
& 27
&\cellcolor[HTML]{e8f4d7}16
&\cellcolor[HTML]{e8f4d7}38          
\\
5
& 15
& 39
& 16
& 38
\\
6
& 14
& 27
& 16
& 38
\\
\bottomrule
\end{tabular}

%% file: JOURNAL/TABLES/Table_compensators.tex
\begin{tabular}{@{}ccccccccc@{}}
\toprule
& \multicolumn{8}{c}{Time constant {[}s{]}}
\\
& $T_1$
& $T_2$
& $T_3$
& $T_4$
& $T_5$
& $T_6$
& $T_7$
& $T_8$
\\
\cmidrule(l){2-9}
$C_P(s)$
& 18.26
& 0.23
& -2.79
& 0.67
& 0.07
& 30
& 
& 
\\
$C_Q(s)$
& 0.27
& 0.57
& 0.27
& 0.57
& 0.27
& 0.57
& 1.28
& 0.13
\\ \bottomrule
\end{tabular}

%% file: JOURNAL/TABLES/Table_parameters.tex
\begin{tabular}{@{}cccccc@{}}
\toprule
Parameter 
& Sn {[}MVA{]} 
& $T_h$ {[}s{]} 
& $T_l$ {[}s{]} 
& $K_P, K_Q$ {[}pu{]} 
& $p_l, q_l$ {[}pu{]}
\\
\midrule
Value
& 50
& 2
& 0.05
& 1.3
& 0.1
\\
\bottomrule
\end{tabular}

%% file: main_file.bbl
\begin{thebibliography}{10}
\expandafter\ifx\csname url\endcsname\relax
  \def\url#1{\texttt{#1}}\fi
\expandafter\ifx\csname urlprefix\endcsname\relax\def\urlprefix{URL }\fi
\expandafter\ifx\csname href\endcsname\relax
  \def\href#1#2{#2} \def\path#1{#1}\fi

\bibitem{kundur1994power}
P.~Kundur, N.~Balu, M.~Lauby, Power system stability and control, EPRI power
  system engineering series, McGraw-Hill, 1994.

\bibitem{IEEE_PSS}
Ieee recommended practice for excitation system models for power system
  stability studies, Tech. rep., IEEE PES (2005).

\bibitem{Shah_2013_1}
R.~Shah, N.~Mithulananthan, R.~Bansal, Oscillatory stability analysis with high
  penetrations of large-scale photovoltaic generation, Energy Conversion and
  Management 65 (2013) 420--429.

\bibitem{IEA_data}
I.~E. Agency, Total primary energy supply (tpes) by source, year and country,
  Tech. rep., International Energy Agency, accessed: 2022-25-05 (2018).

\bibitem{GridCodeREE}
R.~E. de~Espa\~{n}a, Technical norm of supervision on electrical generation
  modules according to EU Regulation 2016/631, (Spanish) (2019).

\bibitem{New_Zeleand_Grid_Code}
T.~N. Zealand, Normal frequency management strategy, Tech. rep., Transpower New
  Zealand (2019).

\bibitem{Larsen1981}
E.~V. Larsen, D.~A. Swann, Applying power system stabilizers part i: General
  concepts, IEEE Trans. on Power Apparatus and Syst. PAS-100~(6) (1981)
  3017--3024.

\bibitem{Liu2016}
Y.~Liu, L.~Zhu, L.~Zhan, J.~R. Gracia, T.~J. King, Y.~Liu, Active power control
  of solar pv generation for large interconnection frequency regulation and
  oscillation damping, International Journal of En. Research 40~(3) (2016)
  353--361.

\bibitem{Sun_2021}
P.~Sun, J.~Yao, Y.~Zhao, X.~Fang, J.~Cao, Stability assessment and damping
  optimization control of multiple grid-connected virtual synchronous
  generators, IEEE Trans. on En. Conversion 36~(4) (2021) 3555--3567.

\bibitem{Basu2021}
M.~Basu, V.~R. Mahindara, J.~Kim, R.~M. Nelms, E.~Muljadi, Comparison of active
  and reactive power oscillation damping with pv plants, IEEE Trans. on Ind.
  App. 57~(3) (2021) 2178--2186.

\bibitem{Rimorov2015}
D.~{Rimorov}, I.~{Kamwa}, G.~{Joos}, Coordinated design of active and reactive
  power modulation auxiliary loops of wind turbine generators for oscillation
  damping in power systems, in: IEEE PES GM, 2015, pp. 1--5.

\bibitem{Knuppel2012}
T.~{Knuppel}, S.~{Kumar}, P.~{Thuring}, M.~{Stottrup}, J.~{Friman}, Towards a
  reactive power oscillation damping controller for wind power plant based on
  full converter wind turbines, in: 2012 IEEE PES GM, 2012, pp. 1--8.

\bibitem{Yongli_2019}
Y.~Zhu, C.~Liu, K.~Sun, D.~Shi, Z.~Wang, Optimization of battery energy storage
  to improve power system oscillation damping, IEEE Trans. on Sust. Energy
  10~(3) (2019) 1015--1024.

\bibitem{Adjust_POD}
N.~Jankovic, J.~Roldan-Perez, M.~Prodanovic, J.~A. Suul, S.~D’Arco, L.~R.
  Rodriguez, Power oscillation damping method suitable for network
  reconfigurations based on converter interfaced generation and combined use of
  active and reactive powers, Intern. Journal of Elec. Power \& En. Syst. 149
  (2023) 109010.

\bibitem{Chau_2018}
T.~K. Chau, S.~S. Yu, T.~Fernando, H.~H.-C. Iu, M.~Small, A
  load-forecasting-based adaptive parameter optimization strategy of statcom
  using anns for enhancement of lfod in power systems, IEEE Transactions on
  Industrial Informatics 14~(6) (2018) 2463--2472.

\bibitem{Yao_2016}
W.~Yao, L.~Jiang, J.~Fang, J.~Wen, S.~Cheng, Q.~Wu, Adaptive power oscillation
  damping controller of superconducting magnetic energy storage device for
  interarea oscillations in power system, Intern. Journal of Elec. Power \& En.
  Syst. 78 (2016) 555--–562.

\bibitem{Arif_2014}
J.~Arif, S.~Ray, B.~Chaudhuri, Multivariable self-tuning feedback linearization
  controller for power oscillation damping, IEEE Trans. on Control Systems
  Technology 22~(4) (2014) 1519--1526.

\bibitem{Deng_2019}
J.~Deng, J.~Suo, J.~Yang, S.~Peng, F.~Chi, T.~Wang, Adaptive damping control
  strategy of wind integrated power system, Energies 12~(1) (2019).

\bibitem{Zhang_2021}
G.~Zhang, W.~Hu, D.~Cao, Q.~Huang, Z.~Chen, F.~Blaabjerg, A novel deep
  reinforcement learning enabled sparsity promoting adaptive control method to
  improve the stability of power systems with wind energy penetration,
  Renewable Energy 178 (2021) 363--376.

\bibitem{Konara_2016}
A.~I. Konara, U.~D. Annakkage, Robust power system stabilizer design using
  eigenstructure assignment, IEEE Trans. on Power Syst. 31~(3) (2016)
  1845--1853.

\bibitem{Shah_2014}
R.~Shah, N.~Mithulananthan, K.~Y. Lee, Low-order robust damping controller
  design for large-scale pv power plants, in: IEEE PES GM, 2014, pp. 1--5.

\bibitem{Liu_2017}
C.~Liu, G.~Cai, J.~Gao, D.~Yang, Design of nonlinear robust damping controller
  for power oscillations suppressing based on backstepping-fractional order
  sliding mode, Energies 10~(5) (2017).

\bibitem{Yuan_2020}
C.~Yuan, X.~Yan, Multi-objective robust tuning of statcom controller parameters
  for stability enhancement of stochastic wind-penetrated power systems, IET
  GTD 14 (2020) 4805--4814.

\bibitem{IEEE_Benchmark}
IEEE, Benchmark systems for small-signal stability analysis and control, Tech.
  rep., IEEE (2015).

\bibitem{MATLAB_2019}
MATLAB, version 9.7.0 (R2019b), The MathWorks Inc., Natick, Massachusetts,
  2019.

\bibitem{Klein1991}
M.~Klein, G.~Rogers, P.~Kundur, A fundamental study of inter-area oscillations
  in power systems, IEEE Trans. on Power Syst. 6~(3) (1991) 914--921.

\bibitem{Lyons2004}
R.~G. Lyons, Understanding digital signal processing, 3/E, Pearson Education
  India, 2004.

\bibitem{Ogata2010modern}
K.~Ogata, Modern control engineering, Prentice hall, 2010.

\bibitem{ISGT_POD}
N.~Jankovic, J.~Roldan-Perez, M.~Prodanovic, Power oscillation damping using
  converter-interfaced generators under constrained active and reactive powers,
  in: IEEE ISGT Europe, 2021, pp. 1--8.

\end{thebibliography}
